# Spatio-Temporal Analysis of On-Demand Transit: A Case Study of Belleville, Canada

**Irum Sanaullah**[Ψ]
Laboratory of Innovations in Transportation (LiTrans)
Department of Civil Engineering,
Ryerson University, Toronto, ON Canada, M5B 0A1
Email: irum@ryerson.ca

**Nael Alsaleh**[Ψ]
Laboratory of Innovations in Transportation (LiTrans)
Department of Civil Engineering,
Ryerson University, Toronto, ON Canada, M5B 0A1
Email: nael.alsaleh@ryerson.ca

**Shadi Djavadian**
Greenfield Labs,
Ford Motor Company,
Palo Alto, California
Email: sdjavadi@ford.com

**Bilal Farooq***
Laboratory of Innovations in Transportation (LiTrans)
Department of Civil Engineering,
Ryerson University, Toronto, ON Canada, M5B 0A1
Email: bilal.farooq@ryerson.ca

*Ψ Equal contribution authors*

*\* Corresponding author*

*Sanaullah, Alsaleh, Djavadian, and Farooq*

## Abstract

The rapid increase in the cyber-physical nature of transportation, availability of GPS data, mobile applications, and effective communication technologies have led to the emergence of On-Demand Transit (ODT) systems. In September 2018, the City of Belleville in Canada started an on-demand public transit pilot project, where the late-night fixed-route (RT 11) was substituted with the ODT providing a real-time ride-hailing service. We present an in-depth analysis of the spatio-temporal demand and supply, level of service, and origin and destination patterns of Belleville ODT users, based on the data collected from September 2018 till May 2019. The independent and combined effects of the demographic characteristics (population density, working-age, and median income) on the ODT trip production and attraction levels were studied using GIS and the K-means machine learning clustering algorithm. The results indicate that ODT trips demand is highest for 11:00 pm-11:45 pm during the weekdays and 8:00 pm-8:30 pm during the weekends. We expect this to be the result of users returning home from work or shopping. Results showed that 39% of the trips were found to have a waiting time of smaller than 15 minutes, while 28% of trips had a waiting time of 15-30 minutes. The dissemination areas with higher population density, lower median income, or higher working-age percentages tend to have higher ODT trip attraction levels, except for the dissemination areas that have highly attractive places like commercial areas. For the sustainable deployment of ODT services, we recommend (a) proactively relocating the empty ODT vehicles near the neighbourhoods with high level of activity, (b) dynamically updating the fleet size and location based on the anticipated changes in the spatio-temporal demand, and (c) using medium occupancy vehicles, like vans or minibuses to ensure high level of service.

**Keywords:** On-Demand Transit, Ride-hailing, Demand Patterns, k-means clustering

## 1. Introduction

The main challenge with the conventional public transportation system in low population density areas is to maintain a reasonably high frequency and low operating cost per trip. During peak hours, the use of buses may be high, but for a significant portion of the day, they may run empty. This affects the efficiency, sustainability and operating cost of the system. Therefore, it is necessary to incorporate advanced design and operations techniques within the traditional fixed-route bus service to make it more efficient and in sync with dynamic user demand. The rapid increase in technology-based transportation, availability of GPS data, mobile applications and effective communication technologies have led to the emergence of the On-Demand Transit (ODT) system (Alemi et al., 2019; Archetti et al., 2018; Djavadian and Chow, 2017). ODT can be defined as a service with flexible routing (origin and destination) and operation (pick-up timing) based on the user's demand. An attractive option is to merge the best of both services to fulfill the total demand i.e., running a fixed service during high demand and ODT in low demand periods. Such a solution has recently been employed by the City of Belleville in the *"Pantonium-Belleville On-Demand Transit pilot project"* that started in September 2018. Pantonium is an optimization company, which provides advanced and innovative technology solutions for public transit. In this pilot, the ODT service replaced a late-night fixed-route, RT 11, which looped around the city. The service is provided for the period of 9:00 pm-12:00 am during the week and 7:30 pm-12:00 am on the weekends. Users can book a trip request through a mobile app named *"on-demand transit" or* via an online portal. However, riders who are not familiar with technology or do not have access to technology can book their trip through a phone call to the customer service number. The passengers who did not use any of the options from the app, online portal, and call centre, can still use the service if they are present at the spot when the bus arrives. Such passengers are called walk-on or ad hoc riders. Bus drivers register them and enter their data into the system. The analysis for walk-on riders was carried out with the rest of the data, except for the waiting time analysis. The data from October to December 2018 indicates that a greater number of trips (71%) were booked through the ODT app as compared to the booking through the call center (7%) and walk-on riders (21%).

For booking a trip, a user provides the information regarding pick-up and drop-off locations, the time for pick-up, the tolerance time, and the number of riders to be picked. After requesting a trip, a user can track the bus location and rider count in the bus. They can also change the pick-up time and location or entirely cancel the trip. Through the driver's application, the bus drivers get directions to navigate through different stops to serve riders and can have ad-hoc stops where needed.

Previous studies related to ODT have either focused on a specific group of users (elderly, disabled and deprived) or investigated very limited factors of supply and/or demand at the individual or zonal levels. Recent related research on ride-hailing (Uber and lyft) services (Alemi et al., 2019; Young and Farber, 2019) are mostly based on the user's survey and simulation data and, therefore, could not provide insight into associated trips analysis for spatio-temporal demand and supply and origin and destination patterns.

Having access to the unique dataset on the operations of Belleville ODT from Sep 2018 till May 2019, we investigated the performance of the service in terms of waiting time (the difference between time requested by a rider and actual bus arrival time), most frequently used stops, and trips demand pattern. Moreover, the DA (Dissemination areas) level zones with higher ODT demand, trip origins and destinations were identified. Firstly, the impact of demographic characteristics (population density, working age, and median income) on the ODT trip production and attraction levels were explored individually and then K-means machine learning clustering algorithm was applied to study the combined effect of these characteristics on the ODT trip attraction levels.

The remainder of the paper is organized as follows: Section 2 presents a background on the studies related to ODT system followed by Section 3 on data description and methodology. Then Section 4 presents the results and discussion on spatio-temporal demand and supply, the impact of demographic characteristics, waiting time and origin and destination patterns. Section 5 presents the policy recommendations and finally, the paper ends with the conclusion and future research.

## 2. Background

On-Demand Transit (ODT) is defined as public transport with a flexible and dynamic routing system based on real-time demand information and auxiliary data (Häme, 2013; Wang et al., 2014). It is considered a transitional mode of public transport mostly provided in low density areas where traditional bus service is not cost-effective and efficient (Papanikolaou et al., 2017; Davison et al., 2014; Koffman, 2004). Previous studies established that the fixed-route bus services are not suitable for low density areas with low demand, particularly fixed routes are expensive and do not provide good service to riders at nighttime. Therefore, ODT is not only the appropriate option to meet the dynamic user demand, but also to provide a higher level of service (Archetti et al., 2018; Papanikolaou et al., 2017; Wang et al., 2014, Edwards and Watkins, 2013; Khattak and Yim, 2004). The potential benefits of ODT include higher efficiency, increased coverage, and convenience with flexible pick-up and drop-off locations, flexible scheduling of timings, increased ridership, better facility for low demand and low-density areas, and enhanced and cost-effective mobility for all users (Khattak and Yim, 2004; Papanikolaou et al., 2017). It can also be used as a solution to the "first/last mile" transit problem and as a feeder to the fixed-route transit (Djavadian and Chow, 2017).

Khattak and Yim (2004) investigated the user's behaviour and perspective related to the Demand-Responsive Transit (DRT) service in the San Francisco Bay Area and evaluated the user's tendency for the DRT system. Respondents were willing to use and even pay more for the DRT system as compared to the traditional bus routes. They gave priority to the factors of reliability, flexible pick-up times, and comfort of pick-up and drop-off stops. Another study on ridesharing (Komanduri et al., 2018) found that riders are willing to pay more for on-demand services.

Li and Quadrifoglio (2010) proposed a methodology to assist the policymakers in designing the applications of the DRT system for feeder transit as compared to conventional fixed-route bus services. Simulation models were developed to calculate the performance of the DRT and fixed-route bus services for different demand levels and service area sizes. It resulted in finding the critical converting points (based on demand densities) between the DRT and fixed-route services. For the performance evaluation, user's walking time, waiting time, and riding time were considered. The performance of the DRT system was found better for low demand rates when higher values were used for the weight of walking time and when the percentage of drop-off was higher. Wang et al. (2014) explored the socioeconomic factors affecting the demand of ODT for the metropolitan region of Greater Manchester. Demand for ODT was found higher in the areas with a low population density, less private vehicle ownership, higher percentage of Caucasian population, and higher level of social scarcity. Wang et al., (2014) stated that very few studies have been conducted to evaluate the ODT demand based on area-wide data, which would result in overlooking the factors that might have a significant impact on the DRT system.

Papanikolaou et al. (2017) conducted a study to explore the critical parameters required for the successful operation of the DRT system and to find out the deficiencies in procedural elements. This study presents a general theoretical assessment framework to evaluate different DRT cases in comparison to traditional public transport based on network and demand characteristics. Recently, Archetti (2018) performed a simulation study to evaluate the ODT system for its potential benefits. The results suggested that the ODT system might be able to fulfill the higher percentage of users' dynamic nature of transportation demand and its efficiency increased with a greater number of users' requests. Haglund et al. (2019) used empirical trip data to evaluate the on-demand micro-transit (vehicle capacity of 9 seats) service with the name "Kutsuplus" in the Helsinki metropolitan region. The results showed that the demand for the service increased with the passage of time and most of the users belonged to the age group between 30 to 44 years. The waiting time was low after the acceptance of a journey offer. However, the average occupancy was low (1.27 persons) with a capacity of 9 seats.

Alemi et al. (2019) investigated the factors affecting the frequency of ride-hailing services (Uber and Lyft) usage in California using data from September till December 2015. The results indicated that the frequency of using ride-hailing increases with the higher activity density and decrease with the increase in mixed land use. While the sociodemographic factors were found to have a significant influence on the adoption of the service, they did not show any impact on usage frequency. Another ride-hailing study was conducted by Young and Farber (2019) for ride-hailing trips in the city of Toronto. The study used the Transportation Tomorrow Survey (TTS) (household travel survey data, 2016) to compare the ride-hailing users' trips and socioeconomic characteristics to that of the users of other modes of travel. They explored the impact ride-hailing has on the level of ridership of other modes of travel in the city. Most of the users of ride-hailing were found to belong to the younger and Millennials generation (20 to 39 years old) and with an annual household income of $100,000 or more. As expected, they concluded that ride-hailing had caused a substantial reduction in the ridership of taxis. Furthermore, they hypothesized that as people will get more familiar with ride-hailing in the future, it might affect the ridership of other modes of travel. They used only 5% of the population, which is limiting in terms of a complete understanding of ride-hailing trip characteristics. Moreover, as TTS represents only one typical day, the temporal analysis was not carried out. A small Canadian town, Innisfil (40,000 residents), has also tried to use Uber (subsidized rides) instead of providing people with public transit, which would have cost almost $1million in 2017. Initially, the plan seemed successful. However, after two years of operations, it is costing the town more ($1.2million in 2019) than running a bus service (Cecco, 2019). In addition, it did not prove to be a sustainable solution as it resulted in a greater number of cars on the road, which raised concerns about greenhouse gas emissions and air quality.

Based on previous studies, it can be concluded that there are supply and demand (area wide as well as individual level) factors, which can affect the operations of ODT (Wang et al., 2014; Brake et al., 2004). The supply side factors include routing, scheduling, booking of trips, types of vehicles, location tracking, communication, fleet size (FS), and level of service (i.e., responsiveness, waiting time). In contrast, the demand side factors include the frequency of users of ODT, the distribution of trips over time, and space and origin and destination patterns. In most of the supply related studies, only limited characteristics of the ODT system have been studied and demand related studies found to be focused only on the specific group of users, for instance old, disabled, and deprived people. In addition, there is a lack of research in temporal and spatial demand analysis for ODT services, particularly macro transit. In this study, detailed supply and demand side factors and their interactions have been investigated. More specifically, an extensive analysis related to spatio-temporal demand and supply, and origin-destination patterns is carried out.

## 3. Data Description and Methodology

This study is based on the unique data sets collected from *Pantonium-Belleville On-Demand Transit pilot project* from September 18, 2018 till May 29, 2019. Belleville is a city located on the Bay of Quinte in Central Ontario and is considered as the centre of the Bay of Quinte region, with a population density of 205 per square kilometer. According to the Census of the year 2016, population of the city has grown to 50,716 residents, which is a 2.6% increase from 2011. There are 22,744 private dwellings with an average household size of 2.3 individuals. Almost 63% of the population belongs to the age group between 15 and 65 years old, which is considered as the middle group of age and shows the population of working age. About 21% of the population is older than 65 years and almost 16% of individuals are younger than 15 years. The median age of the population is 44 years old, which is higher than the national median age of 41 years old (Hershfield, 2014). According to the Census of the year 2016, the median household income (after-taxes) in Belleville is $53,367, as compared to the national average at $61,348. Most of the residents of Belleville are Caucasian (90%) followed by Aboriginal (4%) and South Asian (1.3%). Due to a strong industrial background, the main jobs created in Belleville are related to production, packaging, food processing, supply and transport. Belleville transit service covers ten routes throughout the city, with sixteen buses running seven days a week.

The shift workers in Bellville did not have a high level of service public transit options to travel at night in the industrial park located at the north end of the urban area. To deal with this issue, the night bus service was introduced to connect the industrial park with the city. The fixed night bus service (Route 11) was introduced on January 2, 2018 and operated till September 17, 2018 from 9:30 p.m. until 12:30 a.m. Two 40-foot buses (one in each direction) were used with 30-minute headway, and the fixed-route took one hour to complete. The riders could call to book their trip by scheduling their pick-up and drop-off times and stop. It was a paper-based system and the service was facing the challenge of low ridership (Pantonium, 2019). With the help of Pantonium in September 2018, it was converted to an ODT (Route 11) to increase the efficiency, ridership, and provide service to all the users, particularly shift workers and students to render their mobility easier at night.

The data used in this research include trip, user, stops, and bus GPS data for completed journeys, as shown in Figure 1. Trips data have information about the trip's planned date, origin and destination IDs, and their location (longitude and latitude). Trip origin and destination locations were matched to the stop locations to extract the pick-up and drop-off locations for OD analysis. The most frequently used stops were further analyzed for the origin-destination pattern of their respective trips. User's data contain the information about trip creation and requested time, actual bus arrival time, and the number of riders per request. Information on requested trips was combined with the actual arrival time of the buses to sort out the information on completed and cancelled trips. Furthermore, the UTC timings were converted into local time and average waiting time was calculated for the level of service analysis.

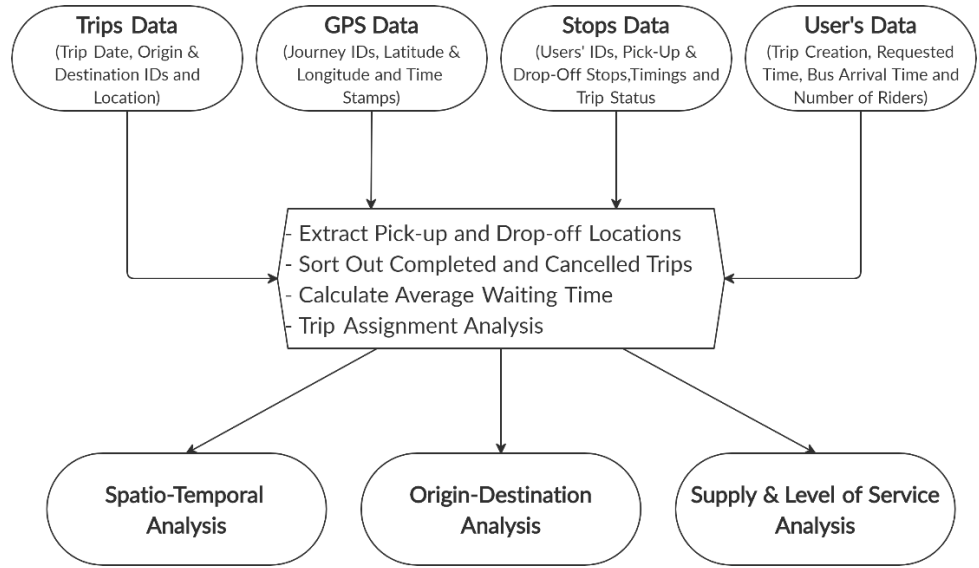

**Figure 1: Steps of analysis**

While stops data include the information about users ID, pick-up and drop-off stops, timings, and the status of the trip (assigned, not assigned, cancelled). The data provide insights into the trip purpose and movement patterns. The GPS data of journeys were projected on DA level zones of Belleville using Geographic Information Systems (GIS). In addition, DA level zones and their demographic data for the city of Belleville was used to explore the demand level analysis. The GPS points of stops were also projected to see the ODT stops locations on a map.

Two different methods were adopted to investigate the impact of Belleville's demographic characteristics on both the ODT trip production and attraction (pick-up and drop-off counts, respectively), as shown in Figure 2. Using GIS, the first method aimed to capture the impact of each characteristic on the ODT trip production and attraction separately. Trip data were used to find the ODT pick-up and drop-off counts and locations. Data from the 2016 census were used to extract the demographic characteristics, including population density, median income, and working age percentages. Three heatmaps were generated using GIS as a base layer to represent the distribution of DA level characteristics over Belleville. The pick-up and drop-off count distributions were overlaid to generate six heatmaps, such that we have two heatmaps per demographic characteristic--one to show its impact on the ODT trip production (pick-up counts) and the other one to show its impact on the ODT trip attraction (drop-off counts).

The combined effects of population density, working age percentages, and median income characteristics on the trip attraction were investigated using k-means machine learning clustering algorithm and elbow method. K-means clustering algorithm, one of the leading algorithms in data mining (Wu et al., 2008), is used to classify unlabelled data points into k predefined clusters, such that the data points that belong to the same cluster are similar to each other, represent the same feature, and differ from the data points that belong to another cluster. Elbow method is a promising technique that can be used to estimate the optimum number of clusters for the k-means algorithm (Bholowalia and Kumar, 2014; Capo et al., 2018; Raval Unnati and Chaita, 2016). At first, k-means algorithm was applied for each of the population density, working age percentages, median income, and drop-off count features, along with the elbow method, to determine the optimum number of clusters for each feature. Thus, each feature is represented by a suitable number of clusters and each cluster contains a certain range of the data, which share similar information. After that, k-means algorithm and elbow method were implemented on all features at the same time. The resulting clusters illustrate the characteristics of the DA that have a certain level of trip attraction in terms of population density, working age percentage, and median income levels.

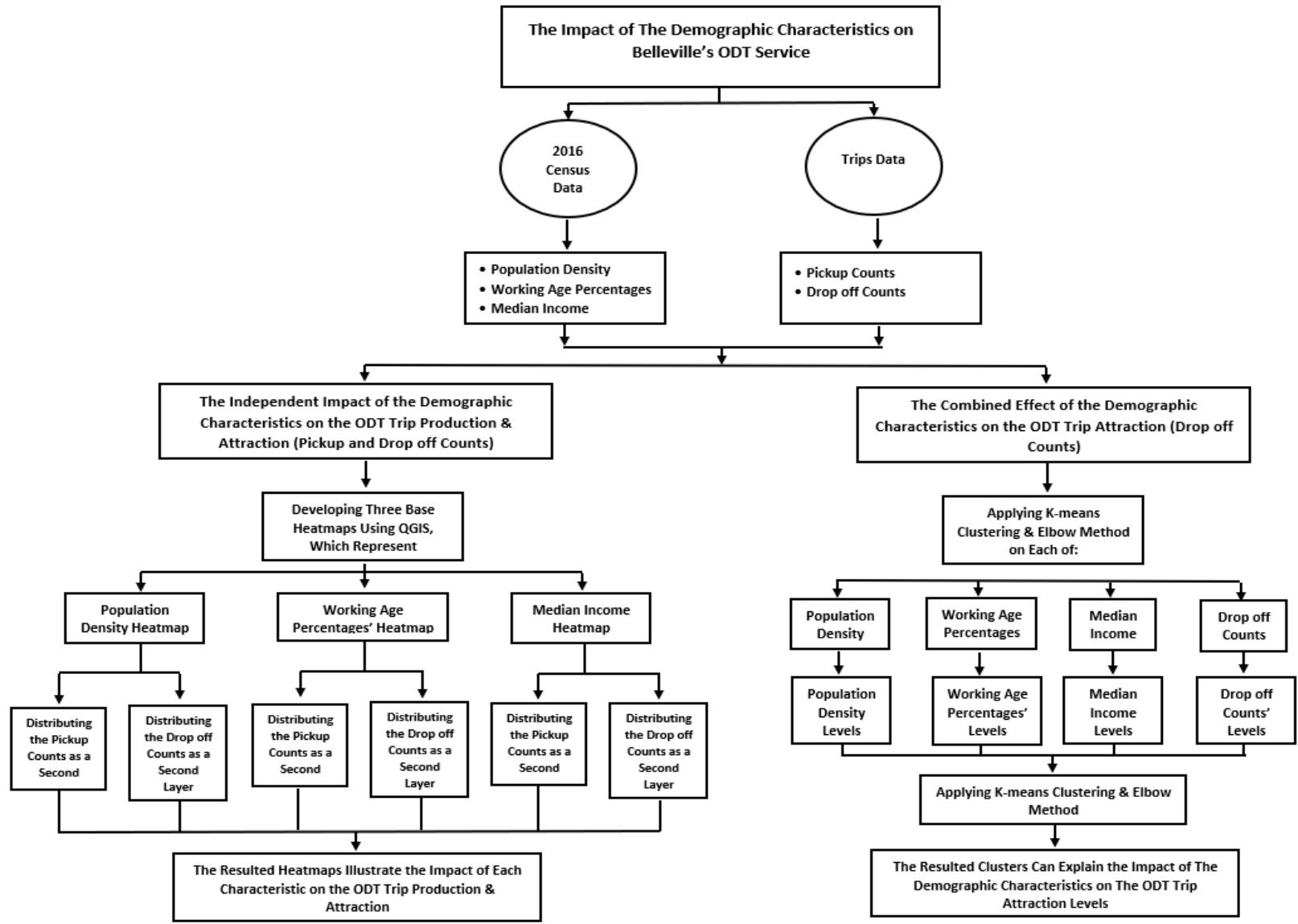

**Figure 2: Studying the impact of demographic characteristics on the ODT trip production and attraction**

## 4. Results

This section first presents the ODT stops, then analyzes the effects of the demographic characteristics on travel patterns of ODT in Belleville as well as the temporal distribution of trips. Supply and demand characteristics are then analyzed in detail. In the end, the origin-destination demand patterns and most frequently used stops are analyzed.

Figure 3(a) represents the Belleville road network, and (b) and (c) show the buffer zones of 100, 200, and 500 metres around the bus stop locations for fixed-route (RT 11) and the pick-up and drop-off stops of ODT bus service, respectively. The stop locations for fixed-route (RT 11) are outside the periphery of the central area, as shown in Figure 3(b). The trip demand is more concentrated in the middle zones of the area, which is clearly indicated from the ODT pick-up and drop-off locations requested by users in those areas and shown in Figure 3(c). The ODT bus service is covering more pick-up and drop-off locations (311) as compared to the conventional fixed-route RT 11 stops (93), which ensures the users' convenience in terms of lesser access and egress distances.

### 4.1 Impact of Demographic Characteristics

As discussed in section 3, the impact of the demographic characteristics on the ODT trip production and attraction levels (pick-up and drop-off count, respectively) was studied in two different ways: (a) the independent impact of the population density, working age percentages, and median income characteristics on the ODT trip production and attraction levels and (b) the combined effect of these characteristics on the ODT trip attraction levels.

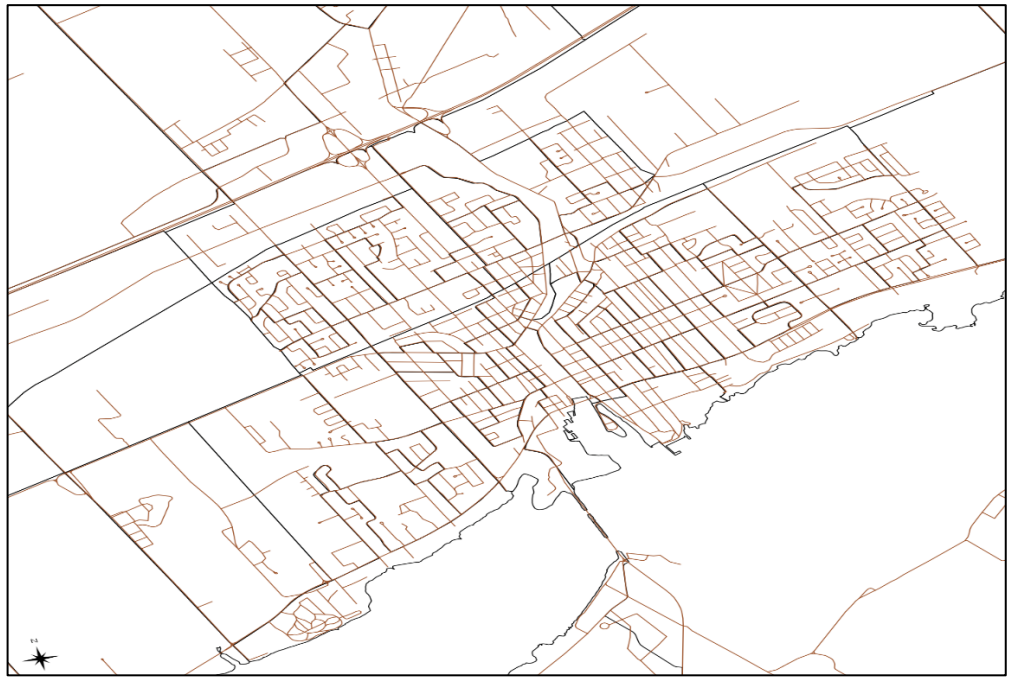

(a) Belleville Road Network

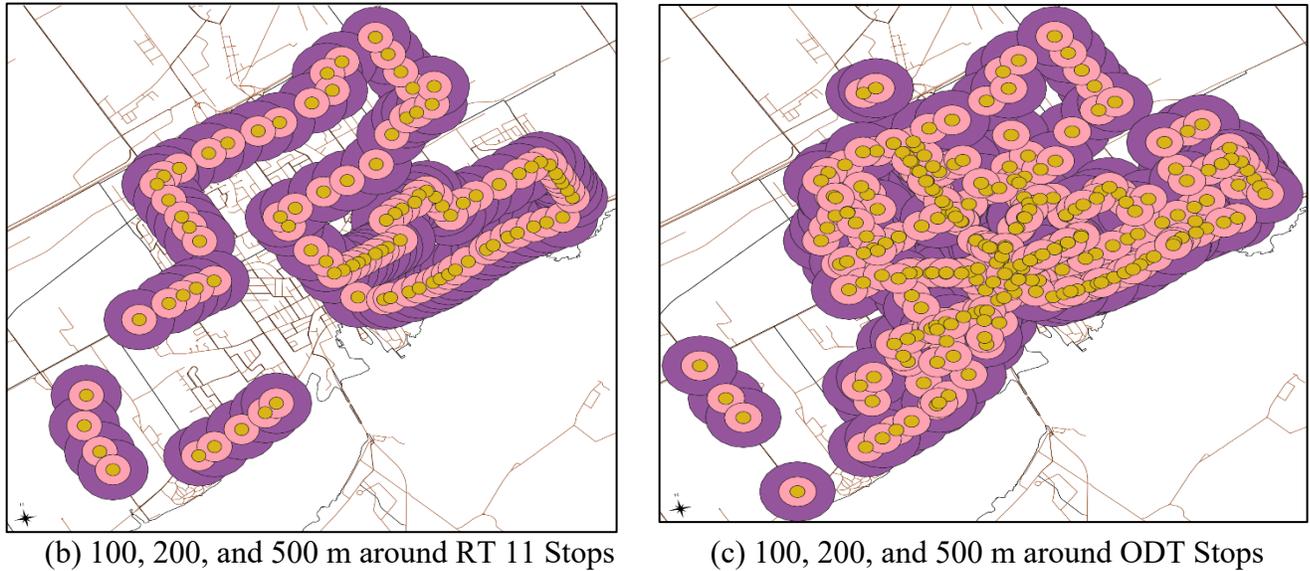

(b) 100, 200, and 500 m around RT 11 Stops    (c) 100, 200, and 500 m around ODT Stops

**Figure 3: Belleville Road network, RT11 and ODT stops overlaid on Dissemination Areas**

### 4.1.1 Independent Impact of Demographic Characteristics

GIS was used to illustrate how each of the population density, working age percentages, and median income characteristics can affect the spatial distribution of the ODT pick-up and drop-off counts. Initially, each characteristic was projected on the DA level zones of Belleville city, which resulted in three base heatmaps. The heatmaps shown in Figure 4 and Figure 5 were developed by projecting the ODT pick-up and drop-off counts, respectively, as second layers on the base heatmaps. Thus, the impact of each characteristic on the ODT trip production and attraction is expressed in a separate heatmap. As Figure 4 depicts, there is no clear pattern between the DAs' demographic characteristics and the ODT trip production levels, since all trip production levels can be found at the same characteristic level. However, the variations in the trip production levels might be explained by the land use type of the DAs. Such that dissemination areas with commercial land use tend to have higher ODT trip production levels than those with residential land use. This trend is observed in Figure 3, where the dissemination areas that have highly attractive places (Labeled pick-up points in Figure 4), like Quinte Mall, Walmart, Metro, Loyalist College, and private companies, have very high levels of trip production, regardless of their demographic characteristics. This trend may imply that most of the ODT users are using the service to return home from work, school, or shopping. In this case, the demographic characteristics have a higher impact on the trip attraction levels than the tip production levels.

On the other hand, it is observed in Figure 5 that the dissemination areas that have higher population density, lower median income, or higher working age percentages tend to have higher ODT trip attraction levels, except the dissemination areas that have highly attractive places (Labeled drop-off points in Figure 5). These dissemination areas tend to have a high trip attraction level, regardless of their population density, working age percentages, or median income values. This trend might be due to the second type of users who are using the ODT service to go to work or do shopping.

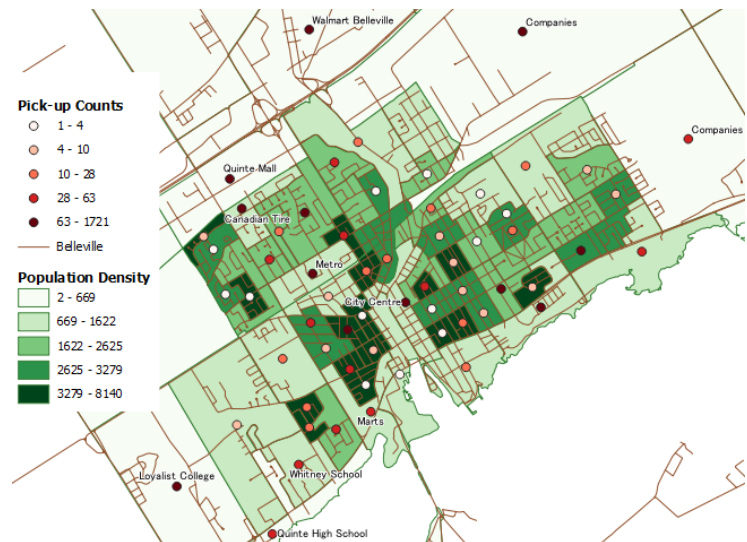

a)  Population density

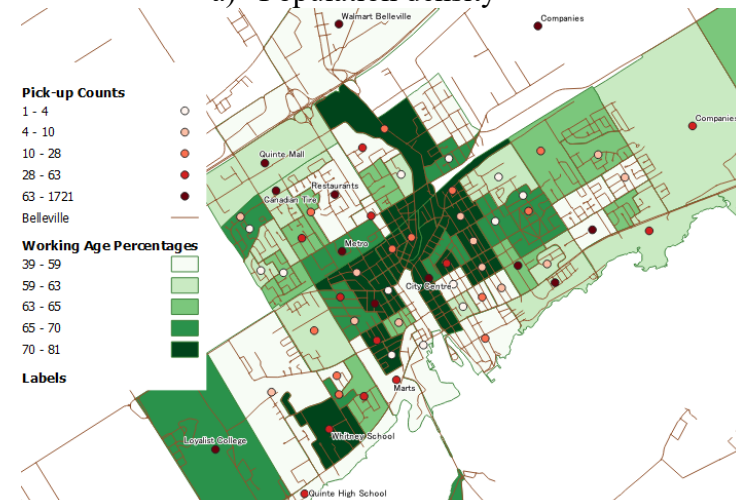

b)  Working age percentages

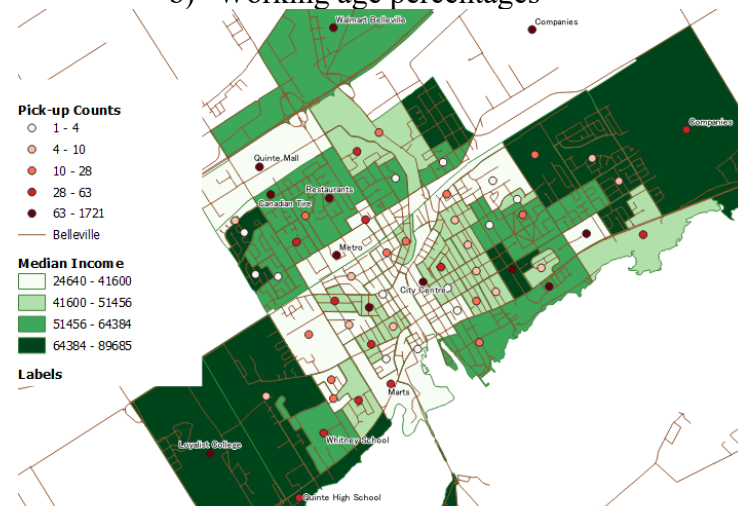

c)  Median income

**Figure 4: The relation between the ODT trip production and a) population density b) working age percentages c) median income**

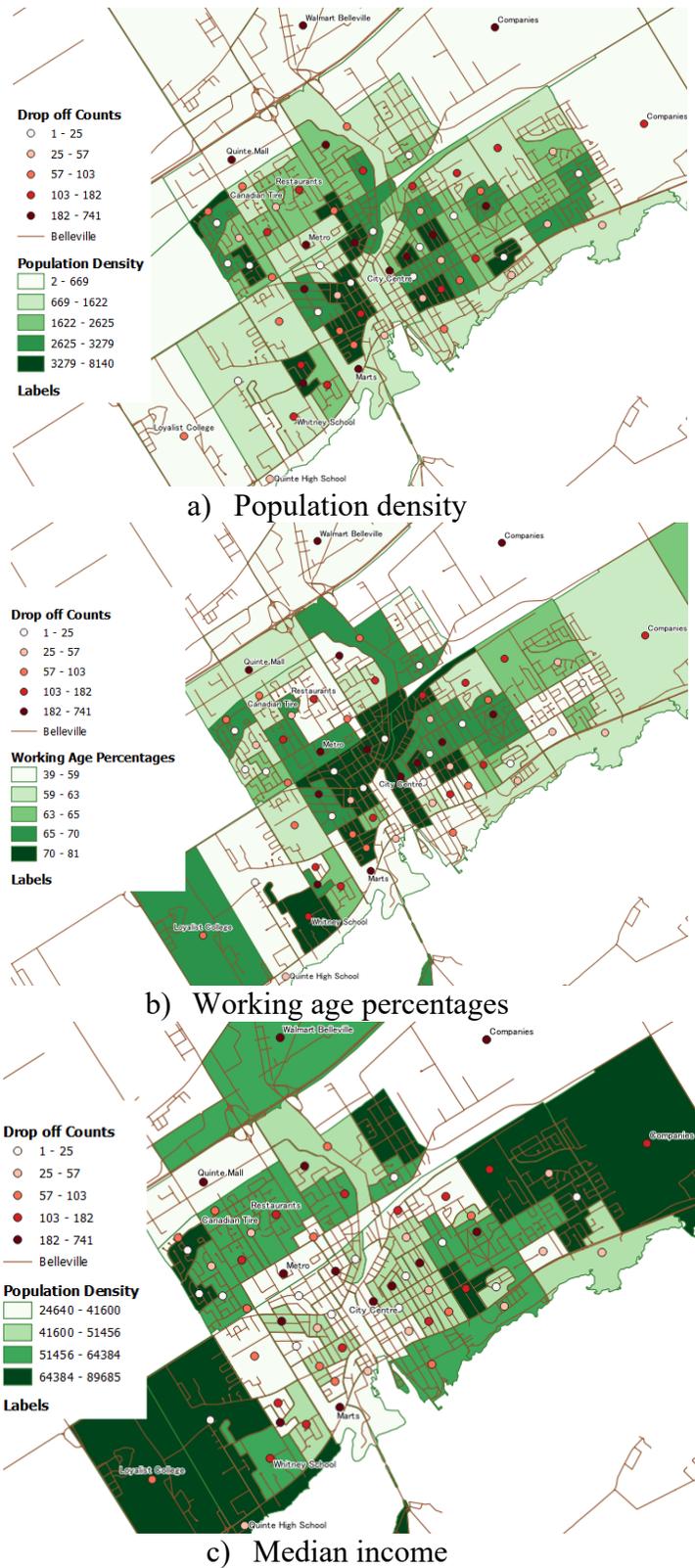

a)  Population density

b)  Working age percentages

c)  Median income

**Figure 5: The relation between the ODT trip attraction and a) population density b) working age percentages c) median income**

**4.1.2 Combined Impact of Demographic Characteristics**

In the previous section, the impact of the population density, working age percentages, and median income on the ODT trip production and attraction levels was investigated independently. In fact, these characteristics are not completely independent of each other, but are instead correlated to various extents. Therefore, it is important to investigate their combined effect on the ODT trip attraction levels. To do so, k-means algorithm was applied along with elbow method on each of the population density, working age percentage, median income, and drop-off count features to classify their values into a suitable number of clusters and the results are shown in Figures 6 and 7.

It is worth mentioning that the elbow method is based on the distortion score, which represents the total within-cluster sum of square distances between each data point and its cluster center. Moreover, the optimum number of clusters is determined through the following steps: 1) running k-means algorithm over a certain range of k-values (number of clusters) and 2) determining the distortion score at each k-value. 3) Then plotting the distortion score as a function of k-values, 4) then selecting the number of clusters so that adding more clusters does not result in a much lower distortion score, which is usually the location of the bend (Bholowalia and Kumar, 2014; Kodinariya and Makwana, 2013; Nanjundan et al., 2019). In light of this information, it is observed from Figure 6(a) through (c) that the bend points occur at k = 4 for population density, working age percentages, and median income features. Since the distortion score decreases rapidly until k=4 and does not decrease significantly after that, this implies that the optimum number of clusters for these features is 4. Similarly, it can be concluded from Figure 6(d) that the ODT drop-off count feature can best be represented by 3 clusters since the distortion score is going down very slowly after this point. Figure 4(a) through (c) shows the optimal clusters representation for the population density, working age percentages, median income, and drop-off counts. Figure 7(a) shows the population density levels, which are: the low density of population (< 1600 persons per sq. km), the average density (1600 – 3500 persons per sq. km), the high density (3500 – 5200 persons per sq. km), and the very high density (> 5200 persons per sq. km). The working age percentages levels are the low percentages of the working age group (< 55%), the average percentages (55% – 63%), the high percentages (63% – 68%), and the very high percentages (> 68%), as shown in Figure 7(b). While the median income levels shown in Figure 7(c) are the very low-income level (< 35,000 CAD), the low income (35,000 – 48,000 CAD), the average income (48,000 – 65,000 CAD), and the high income (> 65,000 CAD). Whereas the ODT drop-off counts' levels are the low-level drop-off counts (< 90 drop-offs), the average level (90 – 200 drop-offs), and the high level (> 200 drop-offs), as shown in Figure 7(d).

In the next step, the feature values were replaced by the results obtained from the first step. This way, there is no need to normalize or standardize the data points before clustering since all features have the same scale. Furthermore, this procedure makes the interpretation of the results easier, as they are directly linked to the levels defined in the previous step. K-means algorithm and elbow method were then applied on all features together as a final step to investigate the relationship between the DA's trip attraction level (drop-off counts level) and its demographic characteristics. The elbow method results presented in Figure 8 indicate that the number of clusters needed to explain the relationship between the population density, working age percentage, median income, and drop-off count levels is 5, as the distortion score goes down slowly after this point.

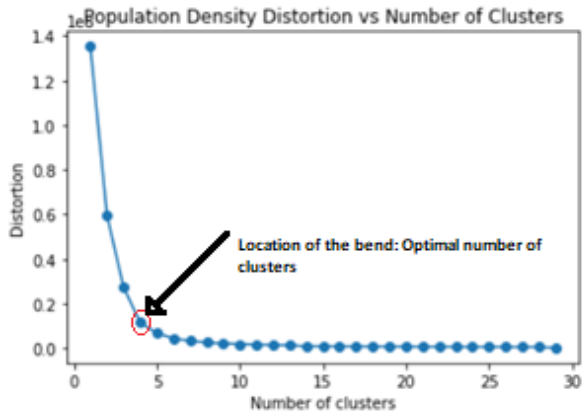

a) population density

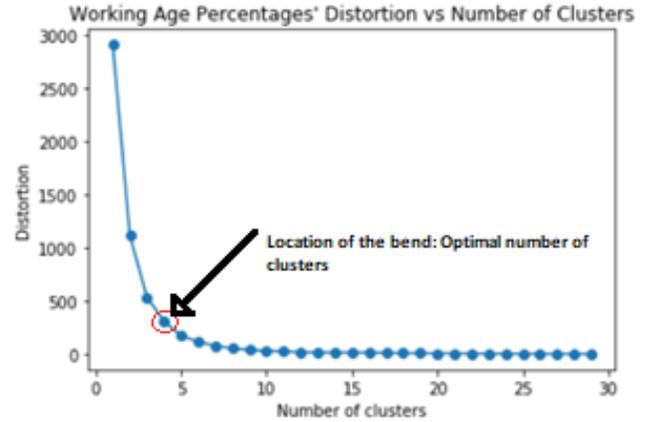

b) working age percentages

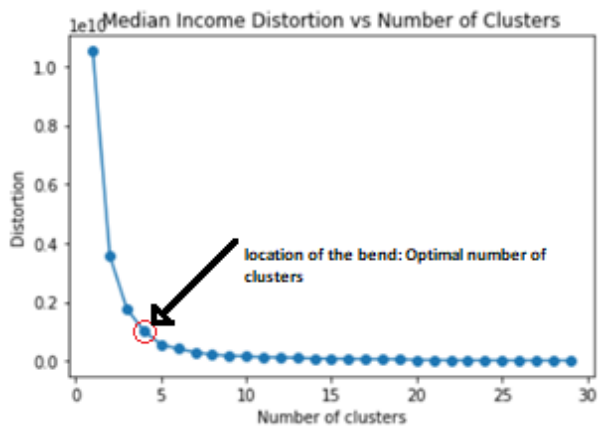

c) median income

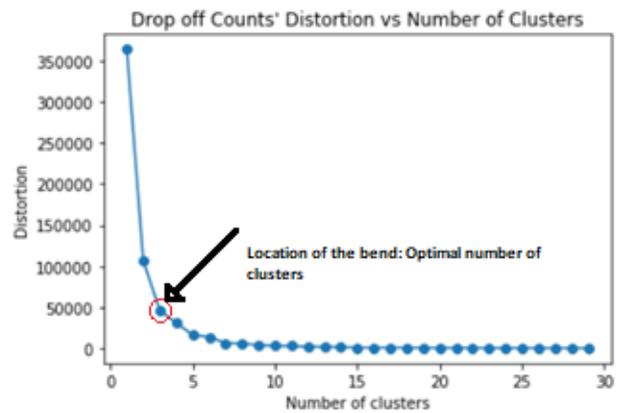

d) drop-off counts

**Figure 6: Elbow method results for a) population density b) working age percentages c) median income d) ODT drop-off counts**

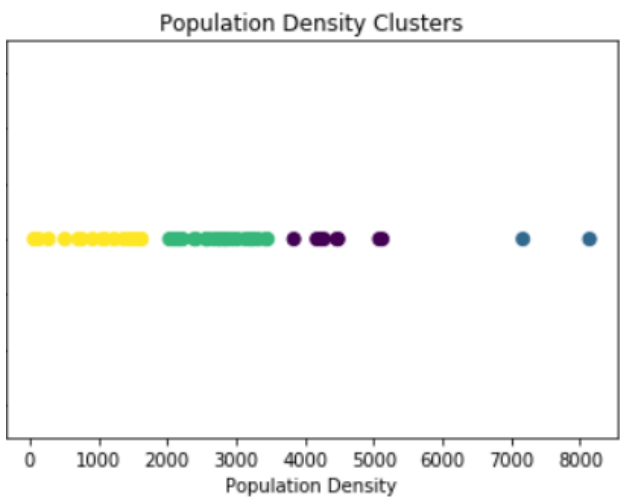

a) population density

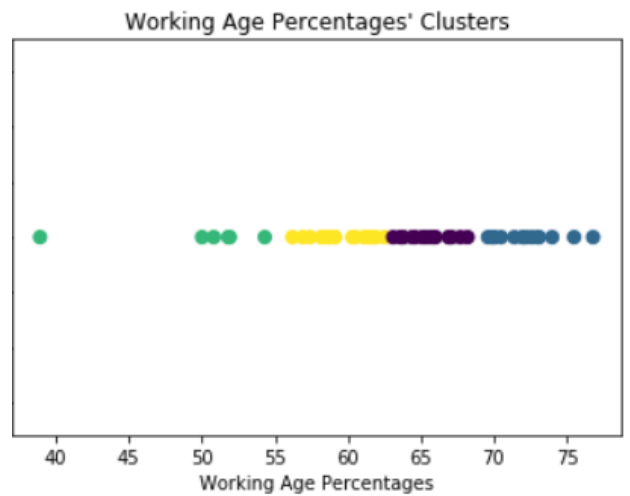

b) working age percentages

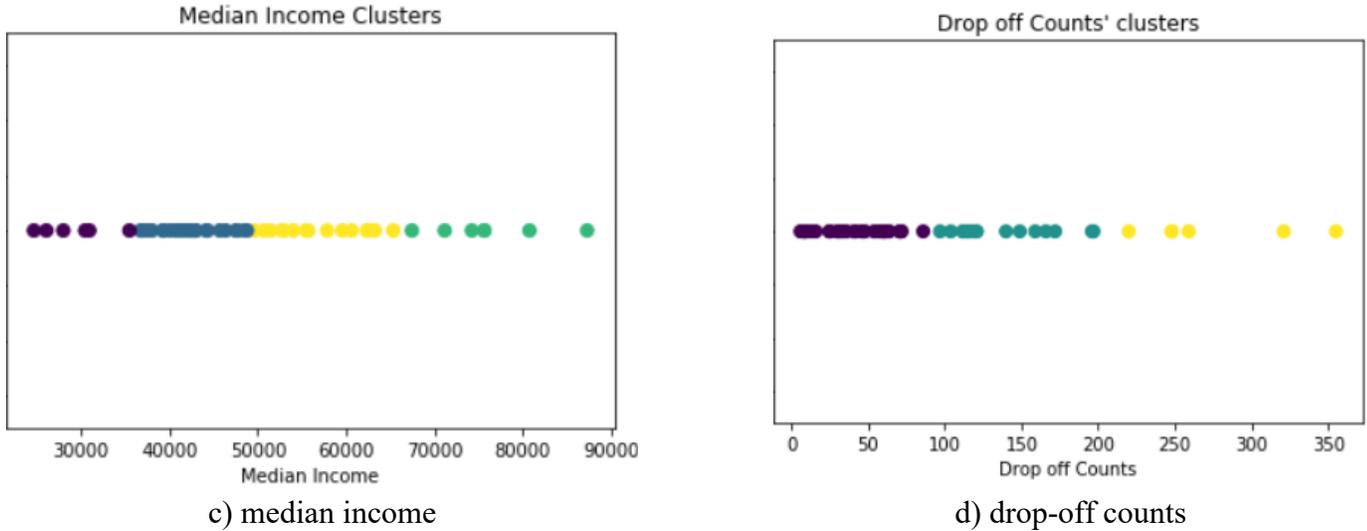

c) median income                                    d) drop-off counts

**Figure 7: Optimal clusters' representation for a) population density b) working age percentages c) median income d) ODT drop-off counts**

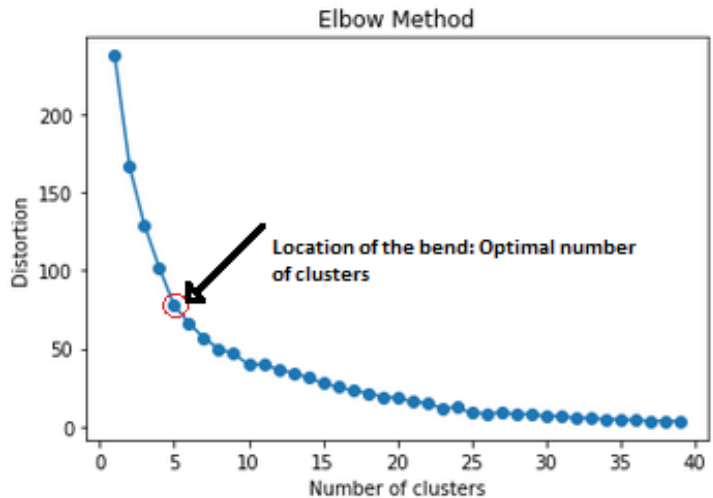

**Figure 8: Elbow method results for the four features**

The flowchart shown in Figure 9 illustrates the results obtained from the clustering analysis. The following points can be drawn from the flowchart (Figure 9):

1. In total, 63 dissemination areas were analyzed to find the general patterns between the trip attraction level they experienced and their characteristics in terms of population density, working age percentage, and median income levels. Among them, 17 dissemination areas have seen a low level of trip attraction, 16 dissemination areas have seen a low to average level, 17 dissemination areas have seen an average level, 11 dissemination areas have seen an average to high level, and 2 dissemination areas have seen a high level of trip attraction.

2. The dissemination areas that have experienced low trip attraction levels are characterized by low to average levels of population density and working age percentages, as well as average to high levels of median income. This pattern is logical since it represents the low-populated areas whose residents have the financial capacity to use their own vehicles or ride-hailing services rather than ODT service to fulfill their needs. In addition, the trip attraction level provides some insights about the land use type of these areas, which is most likely to be residential. The low attraction level confirms the inexistence of shopping centers, large food stores, office buildings, or manufacturing plants that attract the non-residents to do activities like work or shop.

3. The dissemination areas with low to average levels of population density, average to very high levels of working age percentages, and low to average levels of median income characteristics have seen low to average trip attraction levels. Even though these areas represent the low-populated areas as well, they have a higher trip attraction level than those presented in the previous point. This difference could be due to two main reasons: (a) having higher percentages of working age group and (b) lower median income level. Therefore, more people are expected to use the ODT service in these zones to return home from work or shopping. Such zones have a higher working age percentage, but a lower number of residents have the financial capacity to own and use vehicles or ride-hailing services to meet their needs.

4. The dissemination areas that have average to high levels of population density, average to very high levels of working age percentages, and low to average levels of median income characteristics have seen average levels of trip attraction. These areas have a higher level of population density than those presented in the previous point, which resulted in a slightly higher trip attraction level.

5. As expected, the dissemination areas that have high to very high levels of population density, average to very high levels of working age percentage, and very low to low levels of median income characteristics have seen average to high levels of trip attraction. The residents of these poor and densely populated areas have no other option than using the ODT service to return home from work at night due to the financial constraints.

6. Surprisingly, the two dissemination areas that have experienced high levels of trip attraction are characterized by a low density of population, very low to low levels of median income, and high to very high levels of working age percentages. However, this level of trip attraction reveals the type of land use of these poor and sparsely populated areas, which is most likely to be a commercial land use. Quinte Mall, the largest mall in Belleville, is located in one of the dissemination areas, while the other DA covers the city centre with several shopping centers, large food stores, and office buildings. Therefore, the existence of such places increases the demand level of these dissemination areas by attracting more people from other dissemination areas to work and shop.

## 4.2. Trip Distribution over Service Hours

The temporal trip distribution was also carried out for the ODT service, which runs from 9:00 pm to 12:00 am on weekdays and from 7:30 pm to 12:00 am on weekends, as shown in Figure 10. The highest trip demand is observed for the time slot of 11:00 pm to 11:45 pm during the weekdays and 8:00 pm to 8:30 pm during the weekends. The pick-up locations during this time are mostly commercial places and their closing times align with the requested pick-up time. Thus, this pattern may be the result of work-to-home or shop-to-home trips. In the data, most of the top users who have the same origin and destination for their work trips requested to be picked up after 10:30 pm during the weekdays and 8:15 pm on weekends.

## 4.3. Waiting Time Analysis

The waiting time is defined as the difference between the actual arrival time of the bus and the time requested by the rider to be picked up.

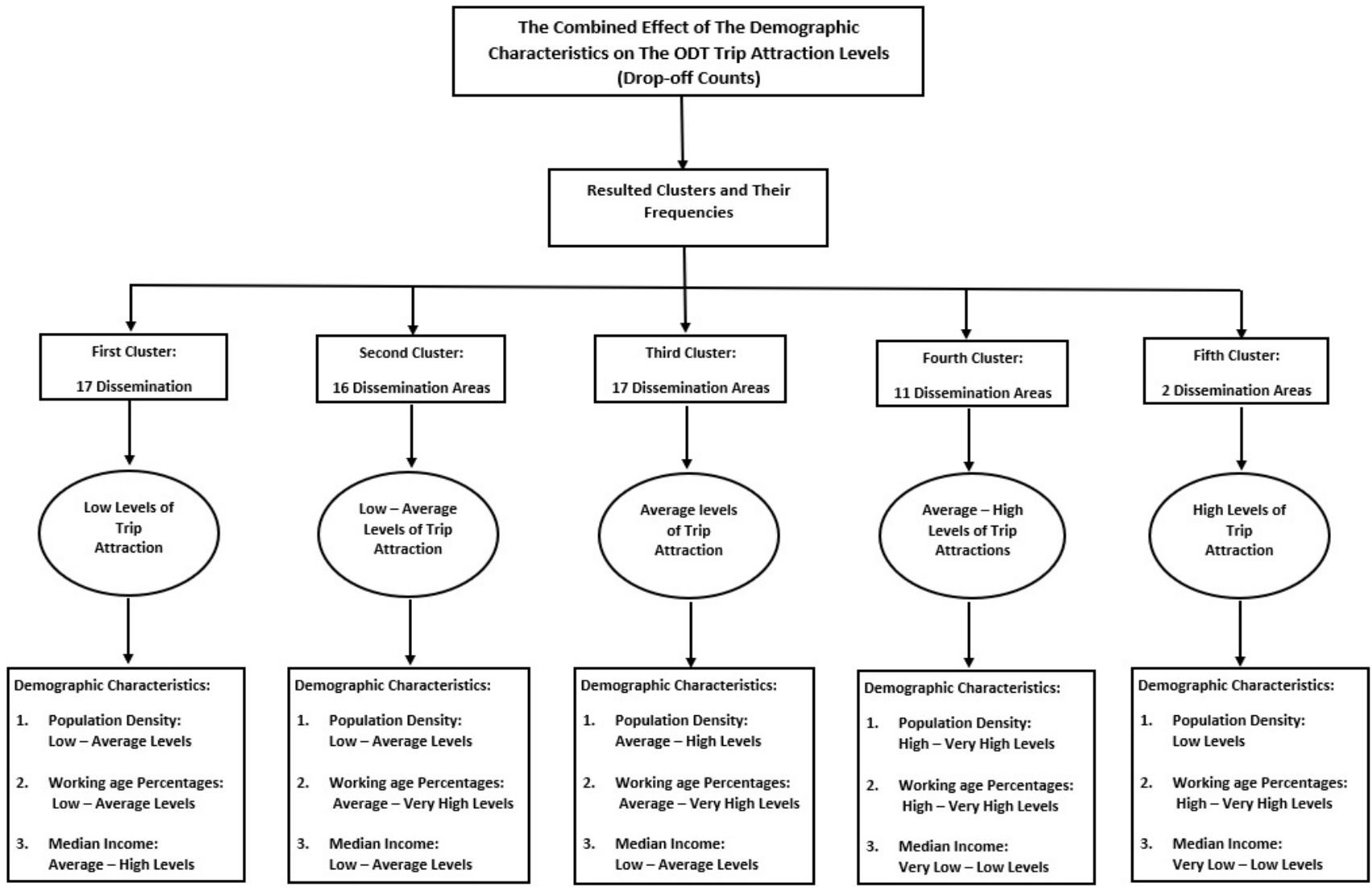

**Figure 9: The combined effect of the demographic characteristics on the ODT trip attraction level**

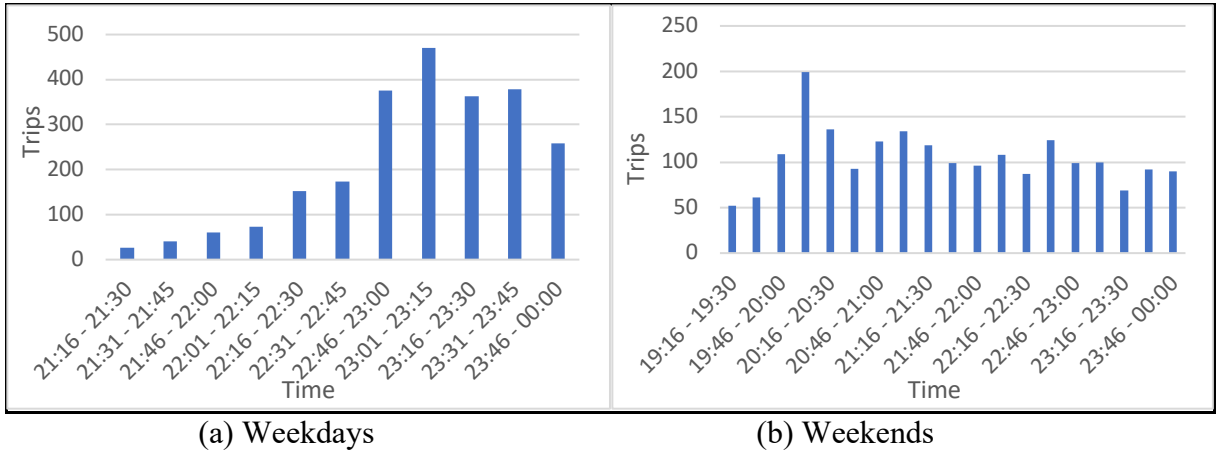

(a) Weekdays          (b) Weekends

**Figure 10: Temporal trip distribution (Sep 2018 -March 2019)**

Waiting time = Actual pick-up time - Requested pick-up time

When riders book their trip through the app, online portal, or call, they mention their requested pick-up time. The ad hoc riders are picked up if they are present at the stop without prior booking. As they did not book the trip earlier and did not request the pick-up time, their waiting time could not be calculated.

Collectively 39% of the trips were found where riders waited for 0-15 minutes, while 28% of the trips had the riders waited for 15-30 minutes (see Figure 11). This is satisfactory in relation to the tolerance time given by the users at the time of booking their trip: 48% of users gave the waiting tolerance time of 15 min and the rest 30 min. The wait time of 30 minutes for 28% of the trips seems rather high. However, the users are attracted to the service due to its pre-scheduling option, flexibility in routing, and convenience in pick-up and drop-off locations. The previous fixed-route (RT 11) late night buses had a frequency of 30 minutes. The main disadvantage of RT 11 was that it did not cover the high demand zones (central area) due to its fixed-route nature.

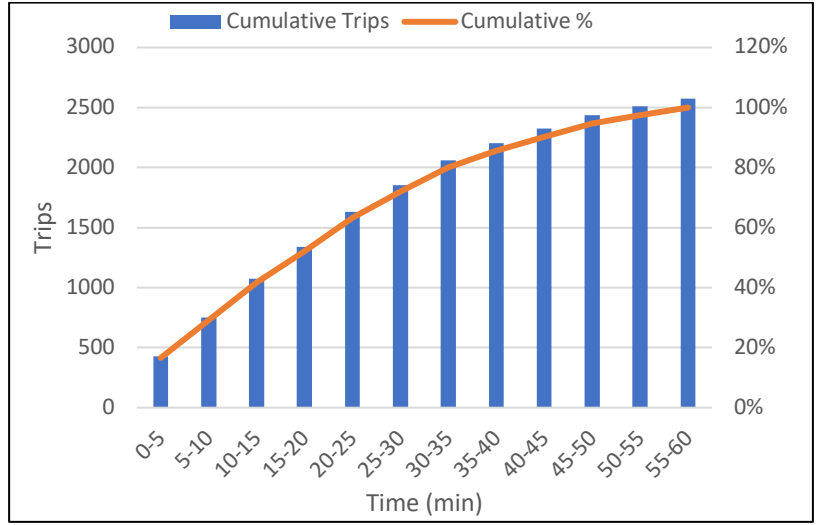

(a) Waiting time and percentage of trips

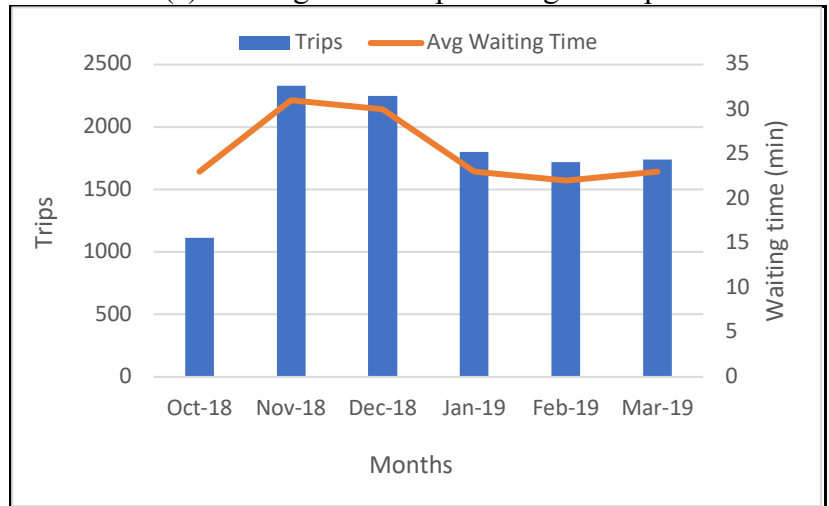

(b) Waiting time and trip demand

**Figure 11: Waiting time analysis (Oct. 2018 -March 2019)**

Figure 11 (b) shows the relationship between the average waiting time and trip demand per month for the data from October 2018 till March 2019. For the first few months (October-December), there is a high variation in the number of trips and the waiting time. However, for most of the users the waiting time (between 21 min to 28 min) is within the given tolerance time. The variation in waiting time is due to the high trip demand in the months of November and December. To fulfill this demand, a fleet size of 4 was used in December. Gradually, the average waiting time and the number of trips became stable in the later months.

## 4.4. Monthly Variations in Supply and Demand

Figure 12 shows the fleet size (FS) of ODT buses for October 2018 till March 2019. It is observed that the fleet size of 3 (number of buses per day) was the most commonly used for these months, except for October, when the fleet size of 1 or 2 was frequently used. Initially, when the ODT service was introduced on 18 September 2018, the fleet size of 1 or 2 was enough to meet the trip demand. However, when the trip demand significantly increased over the holiday season in November and December 2018, fleet size also increased to 3 and 4. For later months (January to March 2019), when the demand became stable, the fleet size started to become stable too.

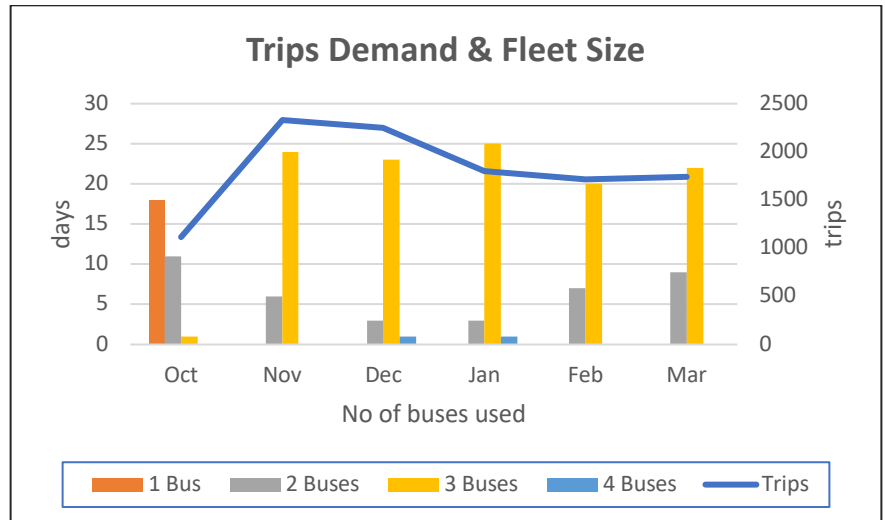

**Figure 12: Demand and fleet size (Oct. 2018 -March 2019)**

### 4.5. Origin-Destination Flow Patterns

Table 1 presents the usage frequency of the ODT service over three months (October to December 2018), as the individual level data were only available for these months. During this time, 2,074 unique individuals used the ODT app, out of which there were 1,420 active and 654 non-active users. The active users are further classified into frequent and infrequent users, based on the classification suggested by Alemi et al. (2019), where the users who used ride-hailing "at least once a month" were considered as frequent users and those who used ride-hailing "less than once a month" were taken as "infrequent users." In this study, the users who used ODT "3 times or more in a month" are considered "frequent users," while individuals who used "less than 3 times in a month" are classified as "infrequent users." As shown in Table 1, the percentage of infrequent users of ODT (62%) is more than the percentage of frequent users (6%). However, the trips made by frequent users are more significant (54%) than the trips completed by infrequent users (46%). To some extent, this is consistent with the study conducted by Alemi et al. (2019), which was conducted to find the factors impacting the frequency of using ride-hailing services (Uber and Lyft) in California.

Figure 13 represents the origin-destination pattern for the 8,302 trips collected from January 2018 till May 2019. For the purpose of origin-destination patterns analysis, 20 most frequently used stops (pick-up and drop-off) were selected during the five months. It shows that Walmart (Supercentre in Belleville) was the most frequently used stop, followed by Lions Community center, Terminal (near to intercity bus terminal), and Jamieson Bone and University (industrial area). Walmart is one of the largest retail superstore chains in Canada, with 411 stores and also the largest employer. It is one of the most used places for buyers and employees. Most of the trips generated at Walmart ended at the stops of College W. and Yeomans N., Sidney street and Moira W., 1 Prince of Wales, Terminal and 179 Palmer stops.

**TABLE 1 Frequency of use of ODT**

| Users/Trips | Frequent Users | Infrequent Users | Non-Active Users | Total |
|---|---|---|---|---|
| **Users** | 134 | 1286 | 654 | 2074 |
| **Users %** | 6% | 62% | 32% | 100% |
| **Trips** | 2584 | 2204 | 0 | 4788 |
| **Trips %** | 54% | 46% | 0% | 100% |

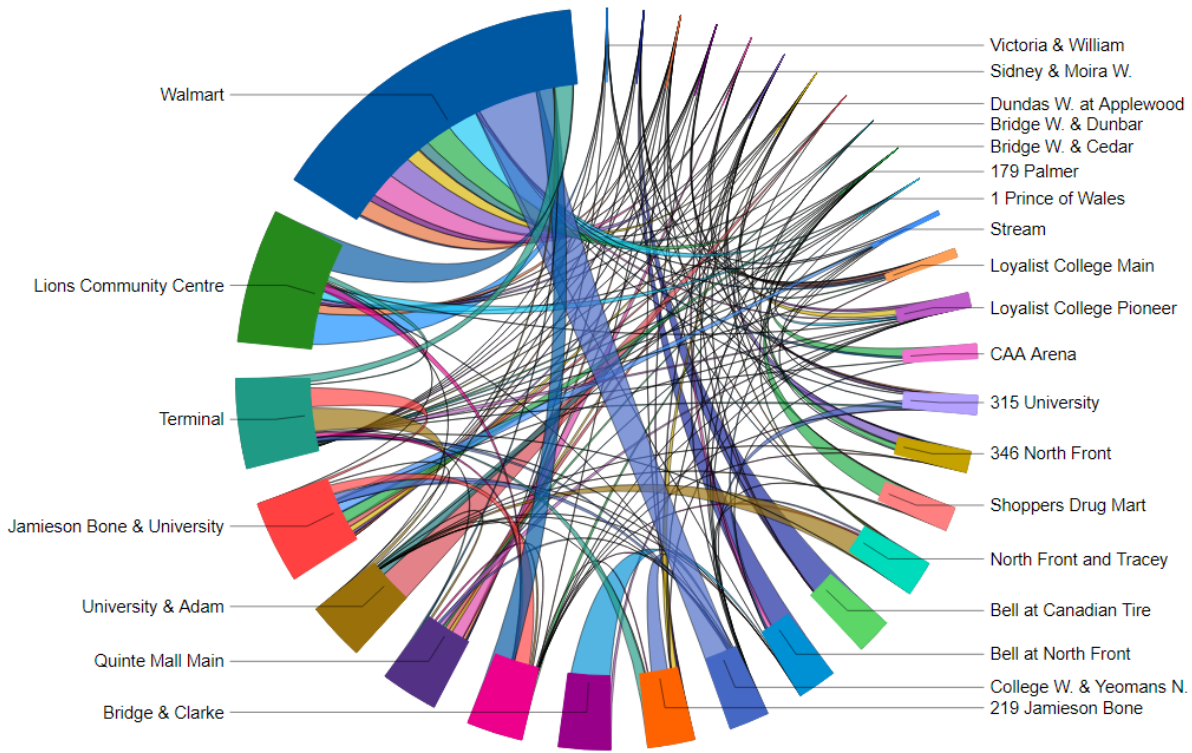

**Figure 13: OD pattern for ODT stop (Jan-May 2019)**

Excluding the Terminal, all other stops are near the zones with single-family dwellings or apartment buildings. This pattern suggests that the trips were made either by workers or shoppers to travel back to their homes. Lions Community centre organizes activities and concerts in the evenings, which leads to the trip generation at this location. The pattern is also apparent in Figure 11, where all the trips that are starting at Bridge and Clarke (residential area) end at the same destination, Bell at North Front (restaurants and shopping area).

Figure 14 shows the ten most frequently used pick-up and drop-off stop locations for ODT service in the Belleville area for the five months duration (Jan-May 2019). From the pick-up point of view, Walmart has the highest usage frequency (20.20%), followed by Terminal (9.40%) and Lions community centre (8.44%). For the drop-off purpose, Terminal (6.55%) is the most popular stop, followed by Walmart (4.96%) and College Street West (4.71%). Walmart Belleville Supercentre is situated on 274 Millennium Pkwy and is a large commercial area with 24 stores and warehouses in surroundings with the opening hours from 7 am to 11 pm. This pattern suggests that individuals working in these stores might be using the ODT service regularly at nighttime for travelling home. Belleville main intercity bus terminal is located at 165 Pinnacle Street. Therefore, commuters could be the frequent users of the ODT service. Most of the Jamieson Bone and University Avenue neighbourhood is an industrial area, which suggests that people working in the area are more likely to use the ODT service at night. There is a residential area near College Street West and Yeomans Street North, which is one of the most frequently used stops for drop-off. The land use surrounding the rest of the top pick-up and drop-off stops include industrial, commercial, residential, educational health and recreation facilities.

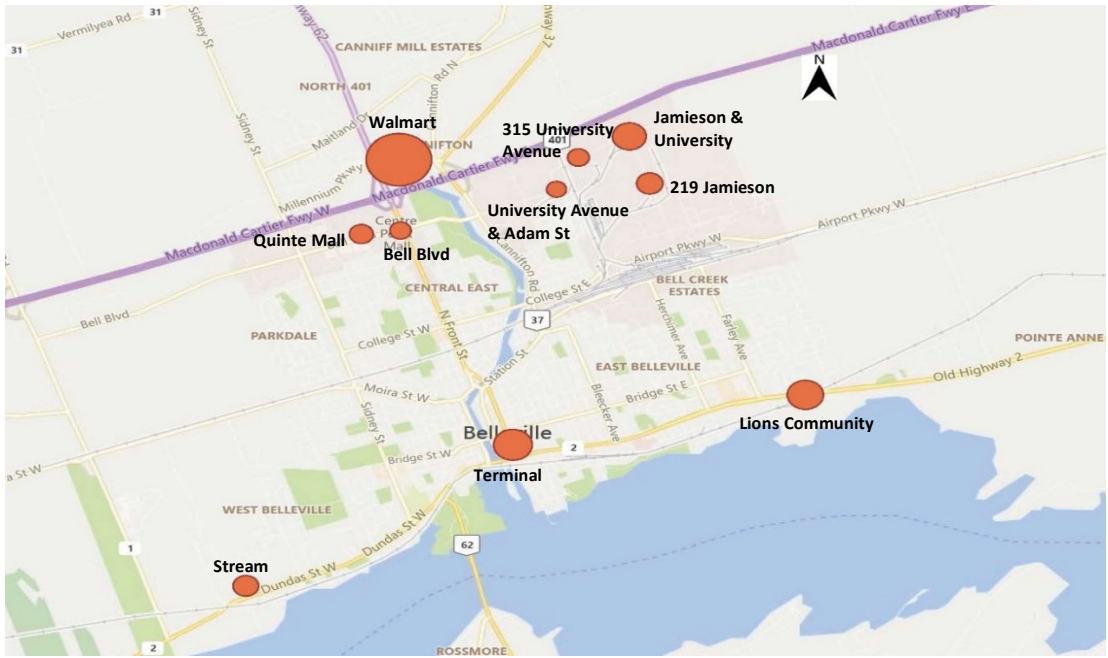

(a) Top Pick-up Stops

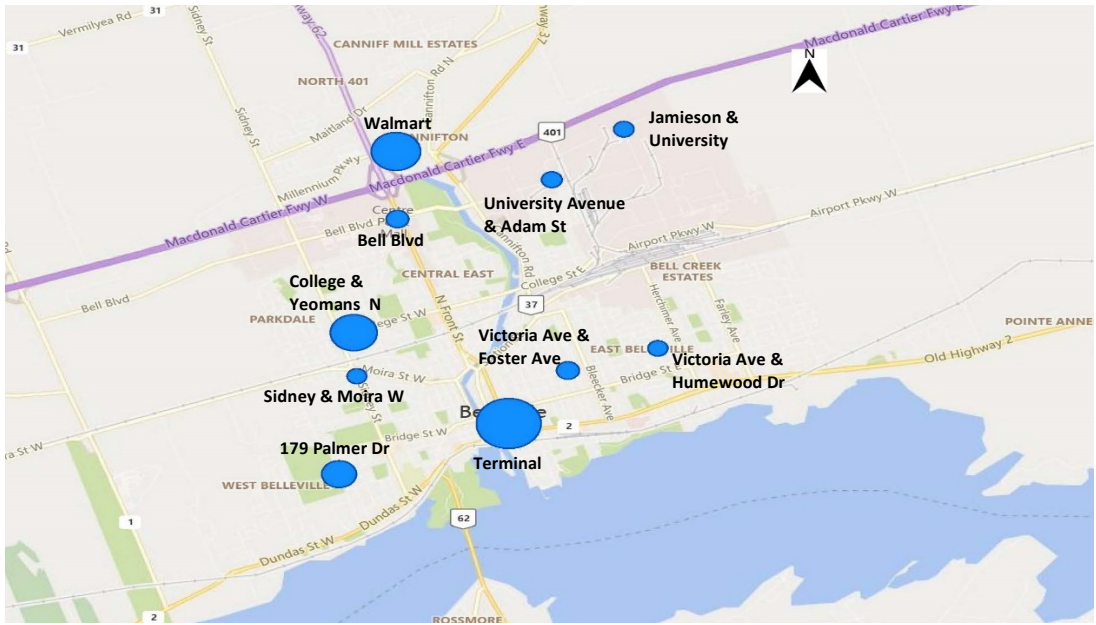

(b) Top Drop-off Stops

**Figure 14: Top 10 Pick-up and Drop-off ODT stops based on usage frequency**

## 5. Discussion and Policy Recommendations

The detailed analysis presented here provides key insights to the municipalities and operators in terms of the planning, design, and operations of new ODT services, resulting in improved sustainability and level of service.

### 5.1. Discussion

This study investigates the impact of spatiotemporal characteristics on the trip attraction and production of ODT night bus service in Belleville, Canada. While comparing the operations of fixed-route bus service from January to March 2018 to the ODT service from January to March 2019, a stable increase of 30% in ridership is observed. Furthermore, Pantonium (2019) reported an increase in the coverage area by 70% and the vehicle mileage reduction of 30%. The fare remained the same for both services, so did the type of bus (40ft). Fixed-route service was using two buses. The ODT service started with the same buses in September 2018. However, when demand increased in the later months, the number of buses was increased to four.

To the best of our knowledge, there are no studies in the existing literature that have specifically investigated the temporal evolution of an ODT service. Furthermore, there is only limited research on investigating the impact of the demographic characteristics on the ODT service trip production and attraction levels. This may be a consequence of having only limited ODT services available on the market. However, this impact was studied on a similar service by Wang et al. (2014), as discussed in section 2. In their study, the authors investigated the correlation between the socio-economic characteristics and the usage-level of demand-responsive transit (DRT) service. The results showed that the areas with lower population density and median income tend to have more trips. It can be noticed that the effect of the population density characteristic on the DRT and the ODT services are diametrically different. The results presented in section 4.1 showed that the dissemination areas with higher population density tend to have higher ODT trip attraction levels. This deviation might be due to the differences in the operating philosophy between the two services. The DRT service was designed to operate mainly in the sparsely populated areas with a low frequency of regular public transit service, explaining the high DRT trip demand levels seen in these areas. In contrast, Belleville's ODT service is designed to be the main public transit service at night by covering all the city's stops. On the other hand, the median income characteristic has the same impact on both services, implying that the users of both services are mainly those who do not have the financial capability to use their own vehicles or ride-hailing services to meet their needs.

Looking at the user's perspective, as discussed in waiting time analysis, despite the waiting time of 15-30 minutes for 28% of trips, riders seem to be satisfied with the ODT service due to its dynamic routing, flexibility in pick-up and drop-off stops, and the larger coverage area of the service. The micro-transit study conducted by Haglund et al. (2019) and Komanduri et al. (2018) revealed a similar finding, indicating user satisfaction with the waiting time in most cases. The study by Khattak and Yim (2004) on the user behaviour related to DRT further support our results. The origin-destination patterns in Figure 13 show that in addition to the frequent users (same O-D trips), ODT is serving 46% of users with very diverse O-D patterns. Our results are further reinforced by Archetti (2018) and Haglund et al. (2019), who suggested that the ODT system is better placed to cater to higher dynamic travel demand.

## 5.2. Policy Recommendations

We analyzed the impact of demographic characteristics and land use in Belleville on the ODT travel patterns, temporal distribution of the ODT trips, and the supply performance. While the analysis is specific to Belleville, several key policy recommendations can still be drawn. These recommendations can potentially be beneficial to the operators and municipalities for sustainable planning, design, and operation of new as well as ongoing ODT projects.

The spatial analysis revealed that the ODT service might have a higher chance of success in the low-density areas containing a mix of neighbourhoods with some commercial/ industrial activity and neighbourhoods with low-income population, multi-family housing, and medium- to low-density residential land use. The results also suggest that during the operating hours, the ODT vehicles should be proactively located (and relocated) near the neighbourhoods with high activity centres that have the highest trip production levels. This strategy will help minimize the users' waiting time and enhance the efficiency of the service. To develop a balance between the performance, cost, and coverage area, agencies should continuously monitor the operations and, if needed, modify the service area of ODT projects by limiting to the areas that have average- to high-level of trip productions or attractions. Spatial distribution of socio-demographics should also be taken into account, such that neighbourhoods with high levels of disadvantaged users (i.e., low-income areas) should be considered in terms of the ODT coverage area.

The temporal distribution of the ODT trips and the supply and demand analyses revealed that public transit usage is expected to see a significant increase in ridership just after converting it to an ODT service. We hypothesize that it is due to the features like the pre-scheduling option, dynamic routing, flexibility in the pick-up and drop-off locations, and larger service area. However, the initial rise in demand can result in higher waiting time values for users in the early days of the service if the same fleet size is used, as was in the fixed-route service. Furthermore, it might discourage a sustained mode shift or induced demand. Therefore, new ODT projects should alter their fleet size according to the actual real-time spatio-temporal demand the system receives, especially at the beginning of the operation, until both the demand and the fleet size become stable. This strategy will help make the service more convenient and competitive to the other travel modes in attracting non-captive users.

The average trip length of an ODT user was 5.3 km, the average journey length of an ODT bus was 2.25 hours, and the average occupancy of an ODT bus was 13.5 passengers. These values imply that the current vehicle type (a 40-ft buses) is not an appropriate type of vehicle for the ODT system, as the operating capacity of a bus is not being exploited to its full extent. Furthermore, a higher occupancy vehicle also deteriorates the level of service, since the in-vehicle time, waiting time, and journey length of the vehicle are significantly affected by the number of detours made. It is expected that the use of a smaller vehicle type, like a van or minibus, for the ODT systems would be more efficient. However, a detailed evaluation of vehicle size in pilots like Belleville ODT or a simulation-based study is urgently needed.

Finally, operators who want to deliver a new ODT service are encouraged to conduct a revealed preference pivoted stated preference survey on the travellers of the proposed areas to examine the suitability of implementing an ODT service, the desirable hours of operation, and the coverage area. In the case of suitability, operators can consider developing a user-friendly app, such that users can easily schedule their trips, modify or cancel scheduled trips, and track the assigned vehicle. There are various third-party platforms currently available that could also be used. Alternative easy and accessible options must be provided for the users who do not have access to the internet or smartphones to allow them to use the service, like booking a trip through a call centre or using the service by just showing up at the stop when the operating vehicle arrives. Moreover, a readily accessible software platform should be designed in such a way that it can anticipate the spatio-temporal demand, proactively relocate the vehicles, manage the trip requests efficiently, assign users the most appropriate vehicle, and continuously update the routing scheme as well as the required fleet size. The resilience of the ODT system should also be analyzed in terms of its recovery from service disruptions and response to emergency situations.

Belleville public transit experienced a sustained and significant increase in its ridership (30%) after moving from fixed to on-demand transit. However, it is not clear if this trend is generalizable to similar size urban areas. Furthermore, it is not clear what are the necessary conditions when it makes sense for urban areas to move from a fixed to on-demand transit. Further research is needed to analyze data from on-demand transit projects in different countries to answer these questions.

## 6. Conclusion

This study is focused on the spatiotemporal analysis of the demand and supply factors of the On-Demand Transit (ODT) provision in the City of Belleville, Ontario, Canada. The ODT service involved converting a fixed-route night-bus to a bus-hailing service with dynamic routes and pick-up/drop-off locations. Various datasets were employed to carry out the spatiotemporal analysis on ODT service, including trips and GPS data over five months (Jan-May 2019), users' data over six months (October 2018-March 2019), and stops data over three months (October 2018-December 2018). The temporal analysis was carried out by analyzing the trip distribution and waiting time estimation over the service hours, variation in supply and demand and origin-destination patterns over months. The spatial analysis was based on the land use and demographic characteristics of the dissemination areas, which is carried out using GIS and the K-means clustering method.

The impacts of demographic characteristics on the ODT trip production and attraction levels were explored in two different methods. The first method was carried out using GIS to investigate the independent effect of population density, working age percentages, and median income on the ODT trip production and attraction levels. The results showed that ODT trip production depends on land use type of the dissemination area rather than its demographic characteristics. The dissemination areas that have highly attractive places have seen high trip production levels, regardless of their demographic characteristics. This pattern suggests that the main purpose of using the ODT service is to return home from work, school, or shopping. On the other hand, the dissemination areas with a higher density of population, a lower median income, or a higher working age percentage have experienced a higher level of ODT trip attraction. Furthermore, dissemination areas with commercial places tend to have a higher level of trip attraction, regardless of their demographic characteristics. The combined effect of the demographic characteristics on the ODT trip attraction levels was investigated using the k-means clustering algorithm along with the elbow method. We classified the dissemination areas into a suitable number of clusters based on their characteristics in such a way that the dissemination areas that belong to the same cluster have similar characteristics and share the same information. The main findings of this analysis method are that the dissemination areas that have low to average levels of population density and working age percentages, as well as average to high levels of median income, have experienced a low ODT trip attraction level. In contrast, the dissemination areas that have seen average to high ODT trip attraction levels are characterized by high to very high levels of population density, average to very high levels of working age percentages, and very low to low levels of median income characteristics. Moreover, the dissemination areas that have experienced high ODT trip attraction levels are those with commercial land-use.

Most of the frequent users of ODT service have the same pattern of movement, travelling for the same origin (commercial) and destination (residential or transit hub) around the same time (between 11:00 pm-11:45 pm), which may suggest that they used the ODT service to return home from work or shopping. This pattern requires better decision making about the fleet size during the hours of operations. The number of drop-off locations was found to be higher than the number of pick-up locations, which indicates that the users used other means of transportation in the morning to go to work and used ODT at night to travel back home. In addition to frequent users (same O-D trips), there is a significant number of trips (46%, as mentioned in the results section) made by infrequent users who travel with diverse O-D patterns. It leads to the conclusion that providing the bus service only to frequent users for the particular time slot will not be a viable solution as it will fail to accommodate the higher percentage of trips demand made by infrequent users. Regarding the stop's usage frequency, it can be concluded that Terminal has been used as a "first-mile solution" by commuters travelling from Belleville centre to the neighbouring location where they may live. The main limitations of this study are the unavailability of individual-level demographics and mobility data, such as passenger's gender and the purpose of the trip. Also, the in-vehicle time cannot be calculated as the drop off times of the users are unknown. This information is useful to investigate the impact of converting the public transit system into ODT service on the users' in-vehicle time. Furthermore, the weather condition data at the time of the trip request would be useful to explain the cancellation of trips.

This work is expected to be insightful for the planning, design, and operations of new as well as ongoing ODT projects. The results related to the impact of demographic characteristics on the ODT trip attraction and production levels suggest that the performance of the ODT system can be further enhanced by (a) starting the operation from the dissemination areas that contain high activity centres and (b) modifying the service area of ODT projects by limiting to the areas that have average- to high-level of trip productions or attractions. The results also suggest that the potential areas for a successful implementation of the ODT system are the areas that contain neighbourhoods with some commercial/industrial activity as well as neighbourhoods with low-income population, multi-family housing, and medium- to low-density residential land use. Moreover, the results of the temporal distribution of the ODT trips and the supply and demand analyses suggest that the fleet size of new ODT projects should be continuously updated according to the actual real-time spatio-temporal demand that the system receives. Furthermore, it is recommended to substitute the current vehicle type of the ODT system with a smaller vehicle type, like a van or minibus, to utilize the full capacity of the operating vehicles and enhance the performance of the system. Using the insights of this work, in the future, we intend to develop a spatio-temporal demand modelling framework for the ODT service. We will explore both conventional discrete choice models as well as data-driven machine learning models. To be able to include the trip purpose and individual user characteristics, we will conduct a detailed revealed preference survey of the existing users. An accompanying stated preference survey will be conducted to understand the effect of different supply and location characteristics in more detail. On the supply side, we will use predictive demand models to optimize fleet size, bus location, and routes dynamically. The developed system will be shared with municipalities so that they can design and operate their own low-cost ODT solutions.

## Acknowledgments

This research is partially funded by the Canadian Urban Transit Research & Innovation Consortium (CUTRIC) and Ryerson University. We are thankful to Pantonium Inc. and the City of Belleville for providing us the ODT operations data that was used in this study. We greatly appreciate Nikki Hera-Farooq's time and effort in helping us improve the quality of this manuscript.